\documentclass[%
 aip,
 amsmath,amssymb,
 reprint
]{revtex4-1}
\usepackage{graphicx}
\usepackage{dcolumn}
\usepackage{bm}

\usepackage[utf8]{inputenc}
\usepackage[T1]{fontenc}
\usepackage{mathptmx}
\usepackage{etoolbox}

\usepackage[separate-uncertainty = true,multi-part-units=single]{siunitx}
\usepackage{tabularx}
\usepackage{multirow}

\makeatletter
\def\@email#1#2{%
 \endgroup
 \patchcmd{\titleblock@produce}
  {\frontmatter@RRAPformat}
  {\frontmatter@RRAPformat{\produce@RRAP{*#1\href{mailto:#2}{#2}}}\frontmatter@RRAPformat}
  {}{}
}
\makeatother

\begin{document}
\preprint{AIP/123-QED}
\title[Short title]{Development of a platform for experimental and computational studies of magnetic and radiative effects on astrophysically-relevant jets at OMEGA}
\author{G. Rigon}
\email{grigon@mit.edu}\affiliation{Plasma Science and Fusion Center, Massachusetts Institute of Technology, Cambridge, Massachusetts 02139, USA}

\author{C. Stoeckl}
\affiliation{Laboratory for Laser Energetics, 250 East River Road, Rochester, New York 14623, USA}

\author{T. M. Johnson}
\affiliation{Plasma Science and Fusion Center, Massachusetts Institute of Technology, Cambridge, Massachusetts 02139, USA}

\author{J. Katz}
\author{J. Peebles}
\affiliation{Laboratory for Laser Energetics, 250 East River Road, Rochester, New York 14623, USA}

\author{C. K. Li}
\affiliation{Plasma Science and Fusion Center, Massachusetts Institute of Technology, Cambridge, Massachusetts 02139, USA}

\date{\today}

\begin{abstract}
    Accurate modeling of astrophysical jets is critical for understanding accretion systems and their impact on the interstellar medium. While astronomical observations can validate models, they have limitations. Controlled laboratory experiments offer a complementary approach for qualitative and quantitative demonstration. Laser experiments offer a complementary approach. This article introduces a new platform on the OMEGA laser facility for high-velocity (\SI{1500}{\km\per\s}), high-aspect-ratio (\num{\sim 36}) jet creation with strong cylindrical symmetry. This platform's capabilities bridge observational gaps, enabling controlled initial conditions and direct measurements of plasma characteristics, crucial for refining astrophysical jet dynamics and improving the models accuracy.\\

    Credit : The following article has been submitted to Physics of Plasma. After it is published, it will be found at https://pubs.aip.org/aip/pop
\end{abstract}

\maketitle

\section{Introduction}
Fast axially collimated outflows, also known as jets, are common astronomical observations \cite{Bally2016, Wu2004}. Such structures originate from most classes of accreting compact objects, ranging from young stellar objects to black holes. On a small scale, these outflows contribute to the dynamics of the system from which they originate, including the removal of some of its angular momentum. Yet, they extend farther than the circumstellar medium and propagate well into the interstellar medium over scales that can exceed the parsec \cite{Frank2014}. On a large scale, these outflows impact the astrophysical dynamics by injecting energy and metallicity far from their object of origin. They contribute to the turbulence of the interstellar medium and even participate in triggering star formation. The interactions of such outflows with their surroundings generate shocks, observable in radio, optical, and x-ray frequencies. Consequently, astrophysical jets are easily observed sources of information regarding the objects from which they originated.

While jets are readily observed, numerous questions pertaining to their formation and subsequent dynamics remain open. It is understood that the nature of these outflows varies widely depending on their system of origin. Jets are mostly composed of ionized matter, some are molecular, others are atomic \cite{Wu2004}, ranging from hydrogen to iron \cite{GarciaLopez2010}, while others consist of relativistic electron-positron pairs (M87) \cite{Lee2019}. Each composition brings its physical subtleties. Although their nature may vary, jets share an apparent stability over long distances and collimation, making them high aspect ratio objects. These seemingly simple features, from which their name originates, have yet to be well understood.

Following the advances in laboratory plasma experiment capabilities, several studies relevant to the physics of jets have been performed. These studies have laid the foundations for astrophysical jet studies scaled to the laboratory. These experiments have shown their usefulness in understanding complex physical processes coupled to jet dynamics and morphology \cite{Albertazzi2014, Li2016, Revet2021, Rigon2022}. For instance, the kink instability, the collimation enforced by magnetic-driven hoop stress and poloidal fields, or the morphology of the jet front and its relationship with instabilities and radiative loss were experimentally studied.

In the present article, we report on the development and initial results of a new platform developed on OMEGA to study high aspect ratio and high-velocity jets. This platform allowed for the creation of a jet in a regime of temperature and velocity not reached by other laser installations. Furthermore, the addition and variation of radiative constraints and magnetic fields should eventually broaden our understanding of jet collimation and their energy balance.

Section \ref{section:setup} describes the setups of both the OMEGA experiment and its related numerical studies. Section \ref{section:result} compares experimental and numerical results obtained from the platform. Section \ref{section:discussion} comments on the limits of the numerical approach and the extrapolation of the experimental results. Section \ref{section:conclusion} provides conclusions.

\section{Experimental and computational platforms\label{section:setup}}
\subsection{Experimental Setup}
\begin{figure}
    \centering
    \includegraphics[width=\linewidth]{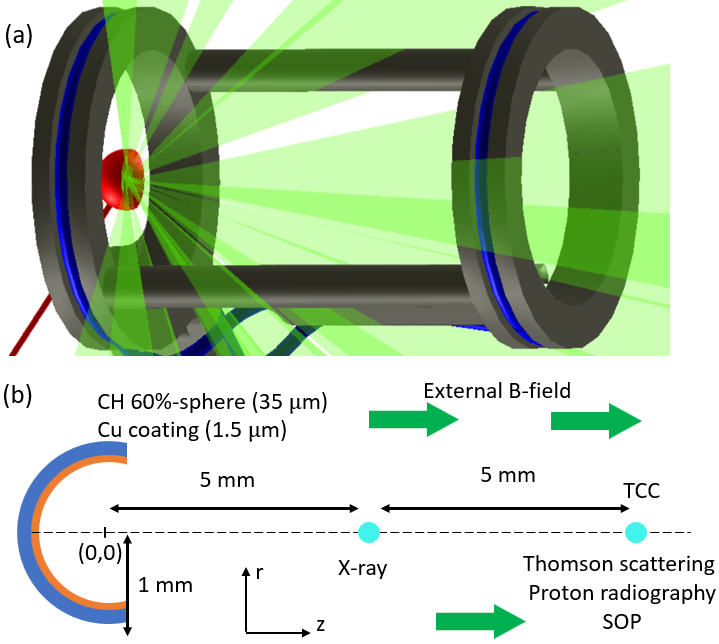}
    \caption{Schematic of the experimental platform. (a) VisRad representation of the main target (red) being driven by 21 lasers (green), and of the surrounding coil in a Helmholtz configuration applying an external magnetic field. (b) Schematic of the main target, a \SI{1}{\mm} radius plastic half-sphere with, in this case, a \SI{1.5}{\um} coating. The field of views' center of the different diagnostics are displayed as cyan dots.}
    \label{fig:setup}
\end{figure}

The experiment was conducted at OMEGA, a multi-kilojoule class laser facility \cite{Soures1996}. Its setup is presented in figure \ref{fig:setup}. Twenty-one laser beams are distributed on the inner surface of the main target, a half-sphere with a \SI{1}{\mm} radius and opened at 60\% of its height. The target is positioned \SI{1}{\cm} away from the center of the target chamber (TCC), with its opening pointing towards TCC.

Each individual laser beam deposits \SI{500}{\J} on a nearly \SI{350}{\um} diameter focal spot in \SI{1}{\ns}, resulting in a laser intensity of approximately \SI{5.2e14}{\W\per\cm\squared}. This process ablates the main target, thus forming 21 plumes of coronal plasma, which expand in the direction normal to the surface of the target. Due to the synchronicity of the laser shot and the sphericity of the target, the plasma plumes collide at the target's center, forming a jet that expands towards TCC. This outcome arises from the conservation of momentum and the pointed location of the focal spot. They are distributed to ensure an equiangular distribution in azimuthal angle for each set of laser beams sharing a polar angle. This experimental setup exhibits near-spherical symmetry around the axis linking the target center to TCC. Henceforth, we refer to the axis of symmetry as the z-axis, the axis orthogonal to it as the r-axis, and denote the origin ($r=0$, $z=0$) at the center of the target and TCC at ($r=0$, $z=$\SI{1}{\cm}).

The subsequent jet expansion is probed using four diagnostics with different focal points. First, an x-ray streak camera coupled to a pinhole observes x-ray self-emission near the main target \cite{xraypap}. This imaging system's line-of-sight is centered on the axis of the jet, \SI{5}{\mm} away from the exit of the main target. It provides a maximal resolution of \SI{\sim 50}{\um} over a nearly 7x\SI{7}{\mm\squared} field of view, accounting for the system geometry, the pixel size and the \SI{15}{\um} pinhole.

The second diagnostic, a streaked optical pyrometry system (SOP) \cite{soppap}, uses an optical system coupled to a streaked camera to observe the emission of the plasma in the visible spectral range. The SOP is centered on TCC with the camera slit nearly parallel to the r-axis, thus allowing measurement of the jet's width.

The third diagnostic is a temporally resolved Thomson scattering (TS) \cite{Froula2011_book}. It employs a 2$\omega$ (\SI{527}{\nm}) laser probe beam, focused on TCC (\SI{35}{\J}, \SI{1}{\ns}). The electron plasma wave (EPW) and the ion acoustic wave (IAW) spectra are recorded by a streaked camera over the \SI{\sim 1}{\ns} duration of the TS probe beam. The investigated volume is estimated to be of the order of 50x50x\SIlist{50}{\um\cubed} surrounding TCC, at the point of interaction between the probe laser and jet.

The last diagnostic is a two-energy proton radiography \cite{PRadpap}. It replaces the TS due to an incompatibility in setups. Protons are produced from the implosion of a \SI{420}{\um} diameter D$^3$He capsule, centered on ($r=$\SI{1}{\cm}, $z=$\SI{1}{\cm}), compressed by 23 laser beams. Given the geometry of the system, the arrival of the proton at TCC is delayed by \SI{\sim 0.92}{\ns} for the \SI{3.02}{\MeV} proton and \SI{\sim 0.69}{\ns} for the \SI{14.7}{\MeV} proton, accounting for their respective velocities and the \SI{\sim 0.5}{\ns} burn duration. After passing through the plasma, the proton distribution is recorded \SI{23.5}{\cm} away from TCC by a stack of CR-39. As the protons are deflected by the electromagnetic fields that they pass through with negligible Coulomb scatterings, this distribution holds information on the field morphology within and surrounding the jet.

One of the objectives of this experimental platform is to investigate the impact of radiative and magnetic effects on the dynamic and morphology of the jets. To do so, both effects are varied during the experiment. The radiative aspect is controlled by altering the composition of the main target, leveraging the tendency of materials with higher atomic numbers (Z) to emit a higher radiative flux \cite{Michel2018}. Thus, two types of main targets are used: a plastic (CH) target for its low Z, and a plastic target coated with a \SI{1.5}{\um} copper layer of higher Z. Simulations indicate that under the employed laser conditions, the copper coating is not entirely ablated, resulting in the jet being composed of a single material.

The magnetic field constraint is externally implemented using a MIFEDS coil \cite{Shapovalov2019}. This coil, built in a Helmholtz geometry, creates a magnetic field nearly parallel to the jet axis. In the center of the system, a maximal field strength of \SI{3.7}{\tesla} is reached after \SI{\sim 0.5}{\micro\s} and stays constant for nearly \SI{1}{\micro\s}. This setup ensures a quasi-constant field from the perspective of the jet dynamic, which is studied for \SI{\sim 10}{\ns}.

\subsection{Computational setup}
To prepare for and to help in the interpretation of the experiment, simulations were performed using the FLASH4 code \cite{FLASH2000,FLASH2014} (v6.2.2 and v7.1). Two-dimensional (2d) radiative ideal magneto-hydrodynamic (MHD) simulations with Biermann-battery terms have been conducted in a cylindrical geometry, with an HLLD-type Riemann solver, a third-order reconstruction scheme (PPM), and a minmod slope limiter. They were performed over a 3x\SI{18}{\mm\squared} domain across 6 blocks, achieving a maximal spatial resolution of \SI{\sim 23}{\um} via a 4th-order adaptive mesh refinement scheme.

The experimental setup is recreated over half its space, given the cylindrical geometry. The lasers are gathered in 3 groups hitting the target at \SIlist{25;45;70}{\degree} angles. Each group delivers the same energy, and total power, as the sum of its beams. Two types of laser configurations are considered in this study: Conf1, only emulates the \SI{25}{\degree} laser group delivering one-third of the total laser energy, and Conf2, employing all three groups for a more realistic energy deposition. Although less realistic, Conf1 allows an easier understanding of the jet morphology due to its lower radial expansion, ensuring the jet remains in the simulation box. 

The target is simulated using a single material either plastic or copper. This approximation holds since the copper coating is not completely ablated according to higher-resolution simulations. The properties of the materials of the target are determined using tabulated opacity calculated by PropacEoS \cite{PROPACEOS} over six energy groups, and tabulated equation of state taken from PropacEoS and SESAME \cite{SESAME} for the plastic and the copper respectively. The chamber's vacuum is represented by a hydrogen gas with \SI{1e-8}{\g\per\cm\cubed} density, with a gamma-law type equation of state and a constant opacity.

\section{Results\label{section:result}}
\subsection{Numerical results}
\begin{figure}
    \centering
    \includegraphics{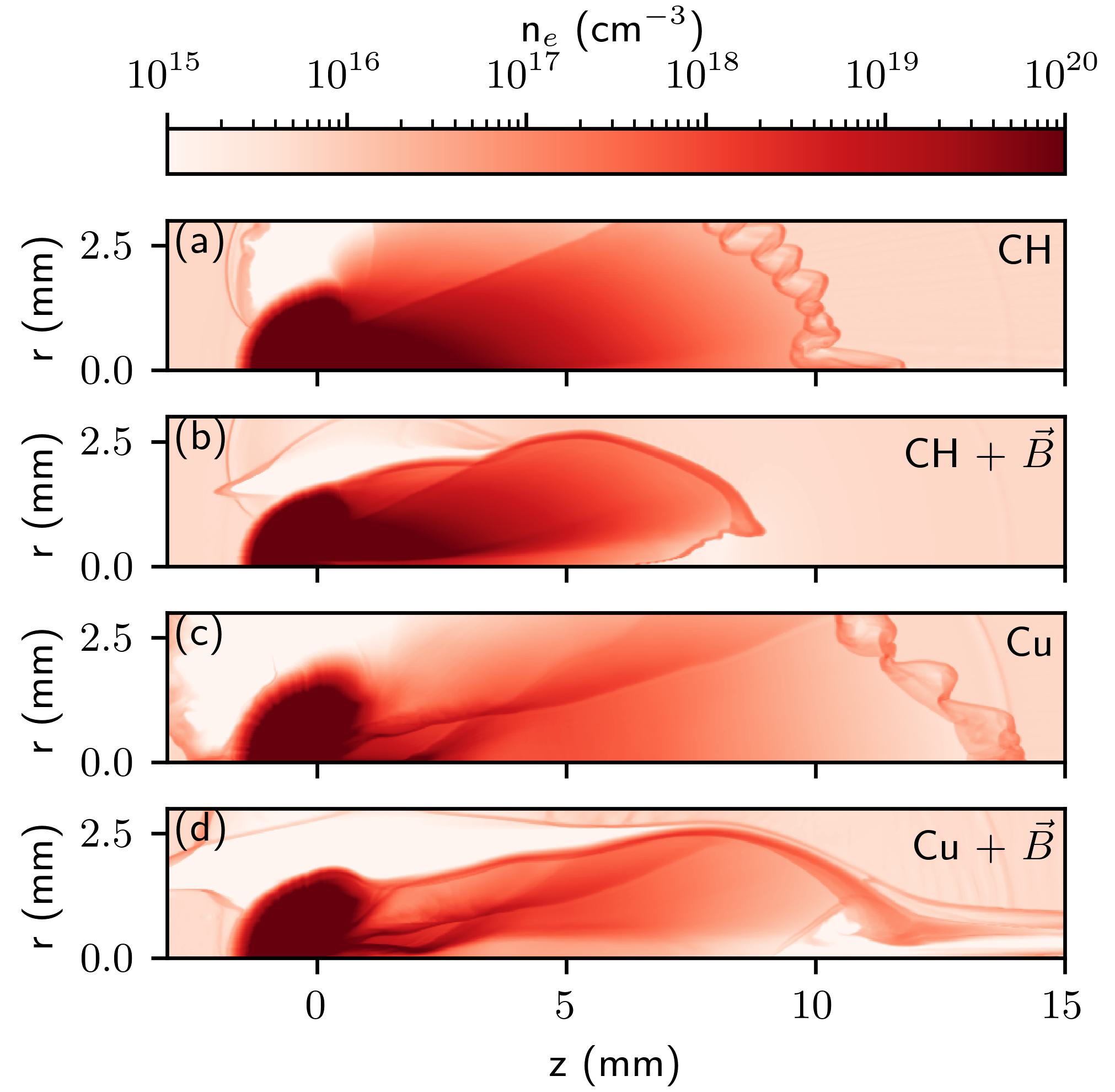}
    \caption{Simulated electron density \SI{5}{\ns} after laser shot (Conf1) for plastic (a,b) and copper (c,d) targets with (b,d) and without (a,c) externally applied magnetic fields. The collimation of the jet is increased in the presence of an external magnetic field and for higher atomic number material. The abscissa is the symmetry axis of the simulation.}
    \label{fig:simulation}
\end{figure}
As the diagnostics employed during the experiment deliver only partial information on the plasma flow, the simulations are necessary to understand the global dynamic of the jet. 

Following the start of the laser shot, the coronal plasma, ablated from the target, expands and reaches the target center in \SI{\sim 0.6}{\ns}. At that point, the radial part of the plasma velocity cancels itself, due to the collision and stagnation of the expanding plasma plumes, and the core of the jet is formed. During this initial stage of the jet formation, the environment surrounding the jet is also affected. The near-vacuum, a tenuous gas, is pushed back by the radiation pressure emitted by the plasma, forming a low-density bubble around the target opening. This bubble is quickly filled by the target's material of low density and temperature. Here, the first difference between copper and plastic targets can be observed. Due to the enhanced radiation produced by the copper plasma, the resulting bubble is nearly three times larger after \SI{1}{\ns} going from \SI{1.2}{\mm} with the plastic target to \SI{3.5}{\mm} with the copper one in Conf1 or from \SI{2.5}{\mm} to \SI{6.5}{\mm} in Conf2. The interaction between the fast-expanding low-density plasma and the limit of the bubble results in the stagnation of the plasma at this interface and the formation of what appears to be a shock. The subsequent creation of oblique shock traveling inward ultimately leads to the deformation of the bubble near the symmetry axis. This results in an elongated shape of the bubble along the symmetry axis in the presence of radiative effects. At later times, this impacts the global morphology of the system.

After the creation of the dense core of the jet surrounded by its low-density shell, the jet propagates on its axis at nearly \SI{1500}{\km\per\s}. At the same time, the jet starts to expand laterally due to thermal effects. As can be observed in figure \ref{fig:simulation}, the rate of lateral expansion is lower in the presence of higher Z material. This is partly a result of the radiative loss, which results in a lower jet's electron and ion temperature (see table \ref{tab:tsmeasure}). As a result of this lower lateral expansion and of the deformation of the bubble at the front of the jet, the copper jet presents a higher aspect ratio (jet length over radius) than the plastic one.

The addition of an external magnetic field also impacts the jet dynamic and morphology, despite the low field value (\SI{3.7}{\tesla}). As can be seen in figure \ref{fig:simulation}, the jet is radially constrained by the presence of the magnetic field, resulting in a smaller radial extension and a higher aspect ratio. As the magnetic field's lines are frozen inside the plasma, they are deformed by the jet lateral expansion. The stretching of these field lines results in a magnetic pressure, which opposes the line deformation and thus the lateral expansion of the plasma. This also impacts the center of the jet, as the lines are squeezed between the expanding coronal plasma during the jet creation phase, leading to the tip of the jet not ending perfectly on the axis. In addition, as shown in figure \ref{fig:simulation}, the presence of the externally applied magnetic field impeded the development of instabilities at the interface between vacuum and jet material.

In summary, the addition of magnetic and radiative effects leads to an increase of the aspect ratio of the jet.

\subsection{Experimental results compared to simulations}

\begin{figure}
    \centering
    \includegraphics{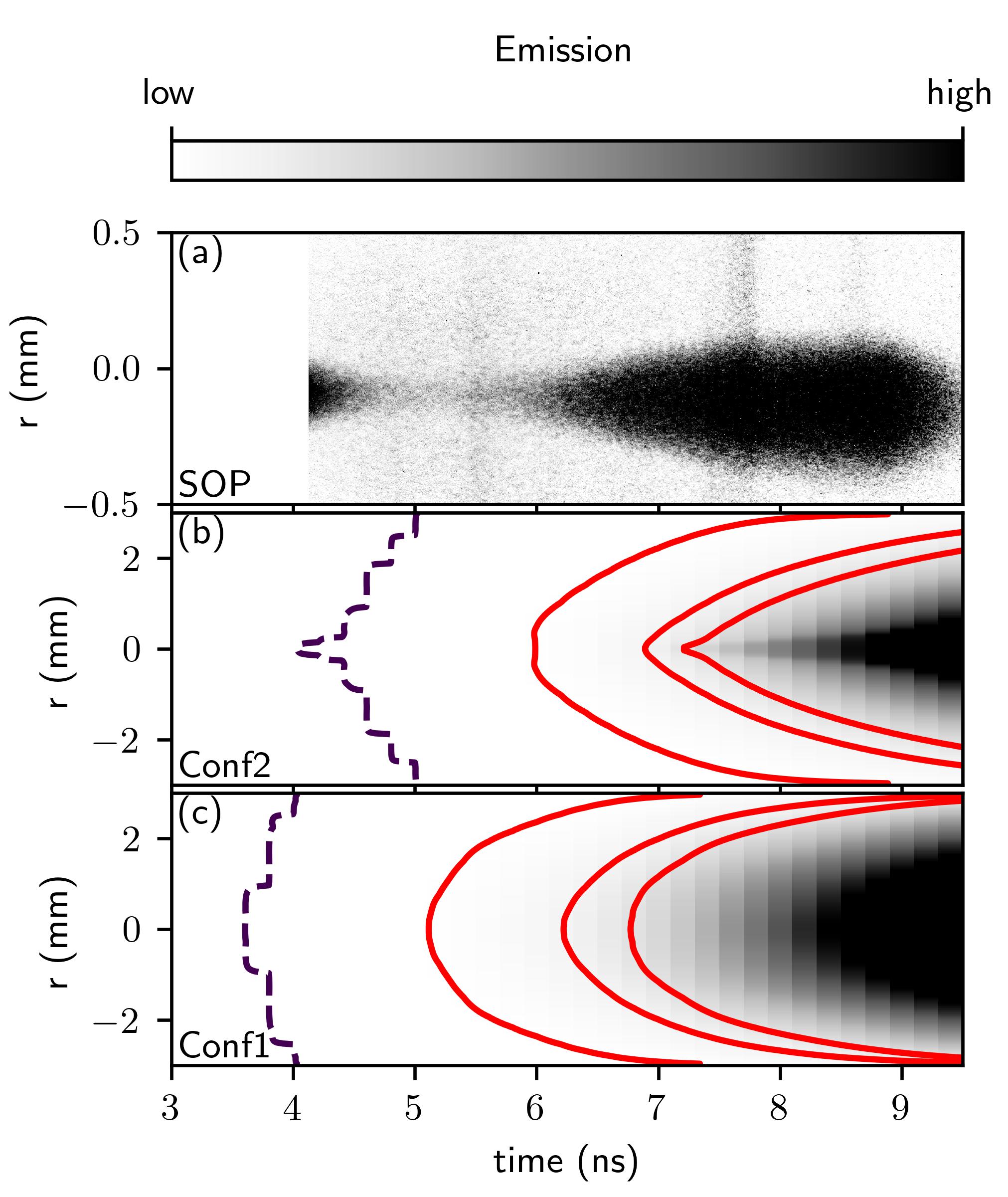}
    \caption{Plasma visible light emission as a function of time both measured with the SOP (a) and simulated Conf1 (b) and Conf2 (c). On (b) and (c) the limit of the plasma outflow is displayed in dashed curved, and the iso-emission contours at \SIlist{1;10;20}{\percent} of the maximum emission are displayed in solid red.}
    \label{fig:sop}
\end{figure}

Experimental results were obtained through a set of eight OMEGA shots. The laser energy delivered to the main target averages to \SI{10.14(23)}{\kJ}, and an average variance in drive laser timing of \SI{0.01}{\ns} in each given shot. Despite the small variation, the set is interpreted as having the same initial conditions (energy deposition profile).

In figure \ref{fig:sop}, the experimental SOP result obtained for the CH jet without B-field is compared to post-process simulations. In the experimental image, the plasma emission is observed after \SI{\sim 6}{\ns} and is \SI{113}{\um} off-center. This marks the arrival of the plasma jet at the TCC plan. The emission becomes larger reaching after \SI{7.5(2)}{\ns} a width of \SI{277(18)}{\um}, which stays approximately constant until the end of the SOP time window. Thus, the emitting part of the jet has an average velocity of \SI{1500}{\km\per\s} and an aspect ratio >\num{36(2)}, given the \SI{1}{cm} distance between the target and measurement plan. One should note the plasma emission is off-centered on the camera strip, this can be understood as an \SI{0.65}{\degree} angle on the target axis.

Synthetic SOPs calculated using a combination of PropacEoS produced Planck emission coefficients, the integrated black body emission over the visible spectrum and an Abel transform, show comparable results. For Conf1, the plasma arrives at TCC after \SI{4.2}{\ns} (dashed curve) and emits a negligible amount of light (\SI{\sim0.01}{\percent} of the calculated maximum). Subsequently the plasma emission increases, first forming a plateau (\SI{\sim0.1}{\percent} of the maximum) until \SI{5.4}{\ns}, before starting an exponential growth until the end of the simulation windows (maximal value). Taking the emission at \SI{8}{\ns} (respectively \SI{9}{\ns}) the spatial distribution consists on the sum of two Gaussians forming a central part with a full width at mid-high (FWMH) of \SI{500}{\um} (respectively\SI{868}{\um}), and a surrounding halo with a FWMH of \SI{2526}{\um} (respectively\SI{2740}{\um}). The halo has a brightness \SI{\sim 40}{\percent} lower than the central part, bringing it to value under the detection threshold of the camera. Thus, the simulation predicts the existence of a broader low-density low-temperature outflow surrounding the core of the jet. However, one should note that contrary to the experimental observation, the width of the jet does not plateau in simulation. Similar results can be observed in Conf2, with slightly earlier arrival of each phase \SI{3.8}{\ns} for the plasma and \SI{4.4}{\ns} for the start of the exponential growth. Furthermore, the halo is slightly larger with a FMHM of \SI{2791}{\um} at \SI{8}{\ns} and \SI{2847}{\um} at \SI{9}{\ns}.

\begin{table}[]
    \centering
    \begin{tabularx}{0.99\linewidth}{|>{\centering\arraybackslash}X|>{\centering\arraybackslash}X|
    >{\centering\arraybackslash}X|>{\centering\arraybackslash}X|>{\centering\arraybackslash}X|}
    \hline
        & & $n_e$ (\SI{e17}{\per\cm\cubed}) & $T_e$ (\SI{}{\eV}) & $v_\text{fluid}$ (\SI{}{\km\per\s})\\
         \hline
         \multirow{2}{*}{CH with B} & Experiment & \num{9(1)} & \num{130(10)} & \num{1530(20)} \\ 
         & FLASH & \num{6(2)} & \num{62(1)} & \num{1541(40)} \\
         \hline
        \multirow{2}{*}{Cu no B} & Experiment & \num{1(1)} & \num{400(80)} & \num{1520(40)} \\ 
         & FLASH & \num{48(21)} & \num{1923(281)} & \num{857(308)} \\
         \hline
         \multirow{2}{*}{Cu with B} & Experiment & \num{3(1)} & \num{130(20)} & \num{1540(30)} \\ 
         & FLASH & \num{1(1)} & \num{16(1)} & \num{1495(43)} \\
         \hline
    \end{tabularx}
    \caption{Results of the Thomson scattering measurement taken \SI{8}{\ns} after laser shot at TCC. Both experimental measurement and simulated values from Conf2 are displayed.}
    \label{tab:tsmeasure}
\end{table}

\begin{figure}
    \centering
    \includegraphics{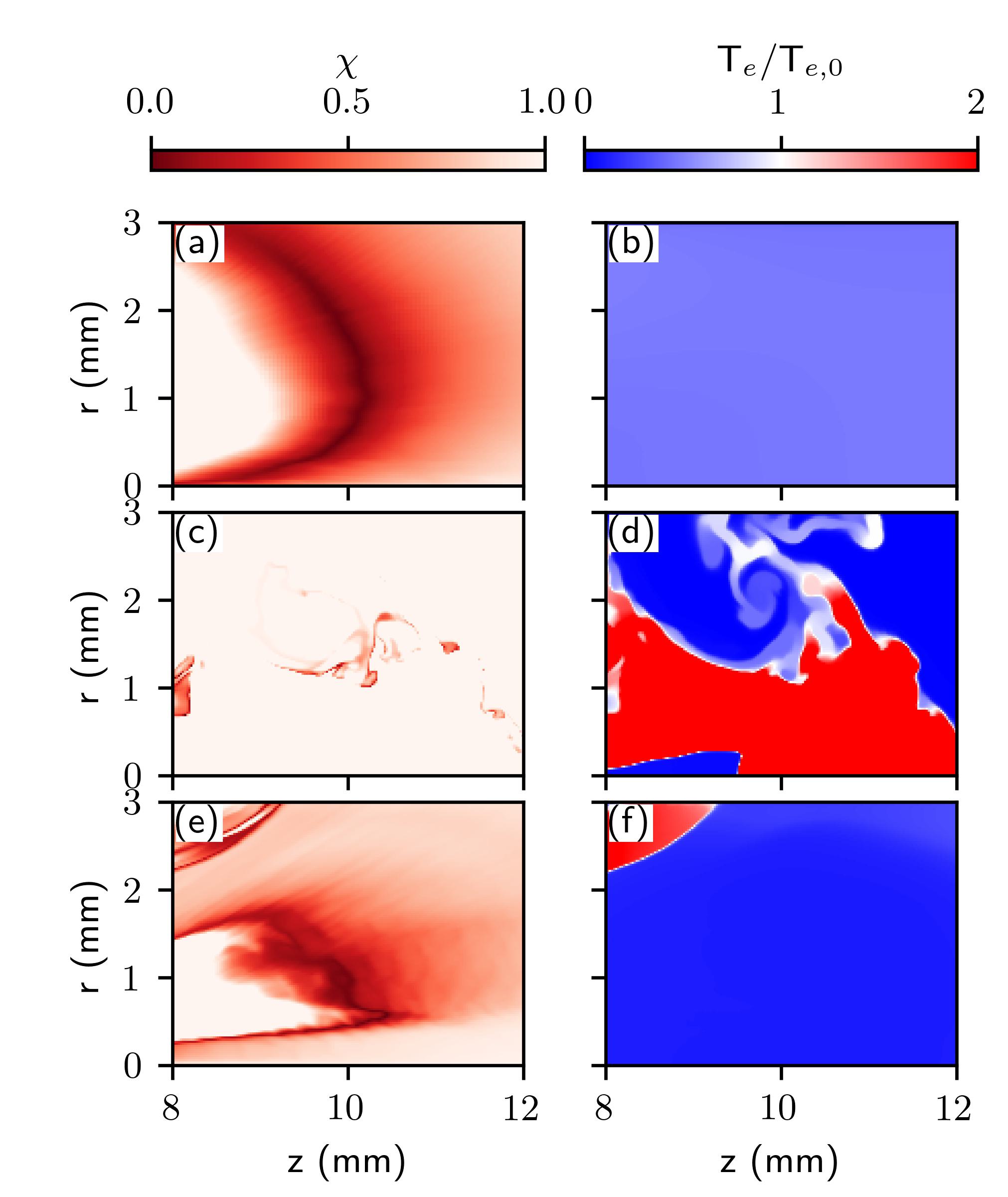}
    \caption{Comparison of the simulated maps (Conf2) to the TS measurement at \SI{8}{\ns} after laser pulse for the CH case with MIFEDS (a, b) and the copper case without (c, d) and with (e, f) MIFEDS in Conf2. On the left, the electron density and fluid velocity combined maps are displayed, with $\chi$ defined in equation \eqref{chidef}. On the right, the simulated electron temperature, $T_e$, is compared to the measured electron temperature, $T_{e,0}$.}
    \label{fig:tsmap}
\end{figure}

The plasma conditions were measured at TCC using TS with a \SI{1}{\ns} probe beam starting \SI{8}{\ns} after the main laser drive. Both EPW and IAW spectra were fitted using a classical TS model for light spectral intensity as defined in \cite{Froula2011_book} and a least-squares method for optimization. The calculation of the spectral density function uses the plasmapy implementation \cite{plasmapy}. The effect of the notch filtering is taken into account based on its spectral response measurement. From the analysis of the EPW, the electron density, $n_e$, and the electron temperature, $T_e$, were retrieved. Based on the IAW Doppler shift, the fluid velocity, $v_\text{fluid}$, was deduced. There is only a small variation of these values (<10 \%) between the beginning and the end of the \SI{1}{\ns} probe beam. The largest variation, \SI{\sim 9}{\percent}, was observed for the fluid velocity in the magnetized copper case, which went from \SI{1570(40)}{\km\per\s} to \SI{1430(50)}{\km\per\s} during this time frame. The fitted values are displayed in table \ref{tab:tsmeasure} with an error given by the optimization method. The fluid velocity around \SI{1530}{\km\per\s} is comparable to the one deduced from the SOP observation.

In comparison, the FLASH values from Conf2 displayed in table \ref{tab:tsmeasure} correspond to the average of the simulated values around the TS sampling region with the standard deviation taken as the error bar. The simulations predicted a lower average electron density and temperature and a comparable fluid velocity. One should note that the sampling volume may contain regions of electron density and temperature that wouldn't be well diagnosed by the employed TS configuration, thus leading to estimates lower than measured. While such phenomena could be responsible for the TS observations, it is worth considering the area surrounding TCC in the hypothesis of a misaligned jet, as suspected from the SOP data. Figure \ref{fig:tsmap} shows a comparison of the simulated physical parameters to their TS-measured equivalent. On the left side of the figure (a, c, e), the electron density, $n_e$, and the fluid axial velocity, $v_z$, are combined through a parameter $\chi$ defined as:
\begin{equation}
    \chi = \sqrt{ \left(\frac{n_e}{n_{e,0}} - 1 \right)^2 + \left(\frac{z_f}{v_{z,0}} - 1 \right)^2 }
    \label{chidef}
\end{equation}
with the subscript $0$ standing for the TS-measured values. $\chi$ can be seen as a type of normalized Euclidean norm, which reaches a minimum of 0 when both electron density and fluid velocity conditions are satisfied. The right side shows a comparison between the simulated and the measured electron temperatures through a simple ratio. In the presence of an externally applied magnetic field (a,e) both electron density and fluid velocity can be retrieved \SI{\sim 450}{\um} off-axis. This could be explained by a \SI{2.6}{\degree} angle of the main target or a similar misalignment of the external magnetic field. Such a small angle, while larger than the one predicted by the SOP, remains in the reasonable range of misalignment of small spherical targets. However, the electron temperature remains too low compared to the observations in both cases (b, f). A possible explanation for this discrepancy between simulations and the experiment would be the heating of the plasma by the Thomson probe beam. Simulations based on the plasma conditions at TCC were conducted to assess such a hypothesis. The electron temperature is indeed raised, going from the initial \SI{16}{\eV} in the magnetized copper case to \SI{61}{\eV} at the end of the probe beam, and the electron density kept nearly constant, going from \SI{1e17}{\per\cm\cubed} to \SI{1.7e17}{\per\cm\cubed} in the same time with a peak at \SI{2.1e17}{\per\cm\cubed}. This effect, while significant, does not fully explain the measured temperature.

The copper case without an external magnetic field (c, d) presents another type of dynamics. Contrary to the cases with magnetic fields, the electron temperature is higher in the simulation. Furthermore, the three parameters can be reached in the same place, on a line that seems to fluctuate around $r\sim$ \SI{1}{\mm}. The Thomson measurement would thus be explained by an angle of \SI{5.7}{\degree}, which seems quite large. In this specific case, the effect of plasma heating by the Thomson beam is negligible according to simulations.

\begin{figure}
    \centering
    \includegraphics{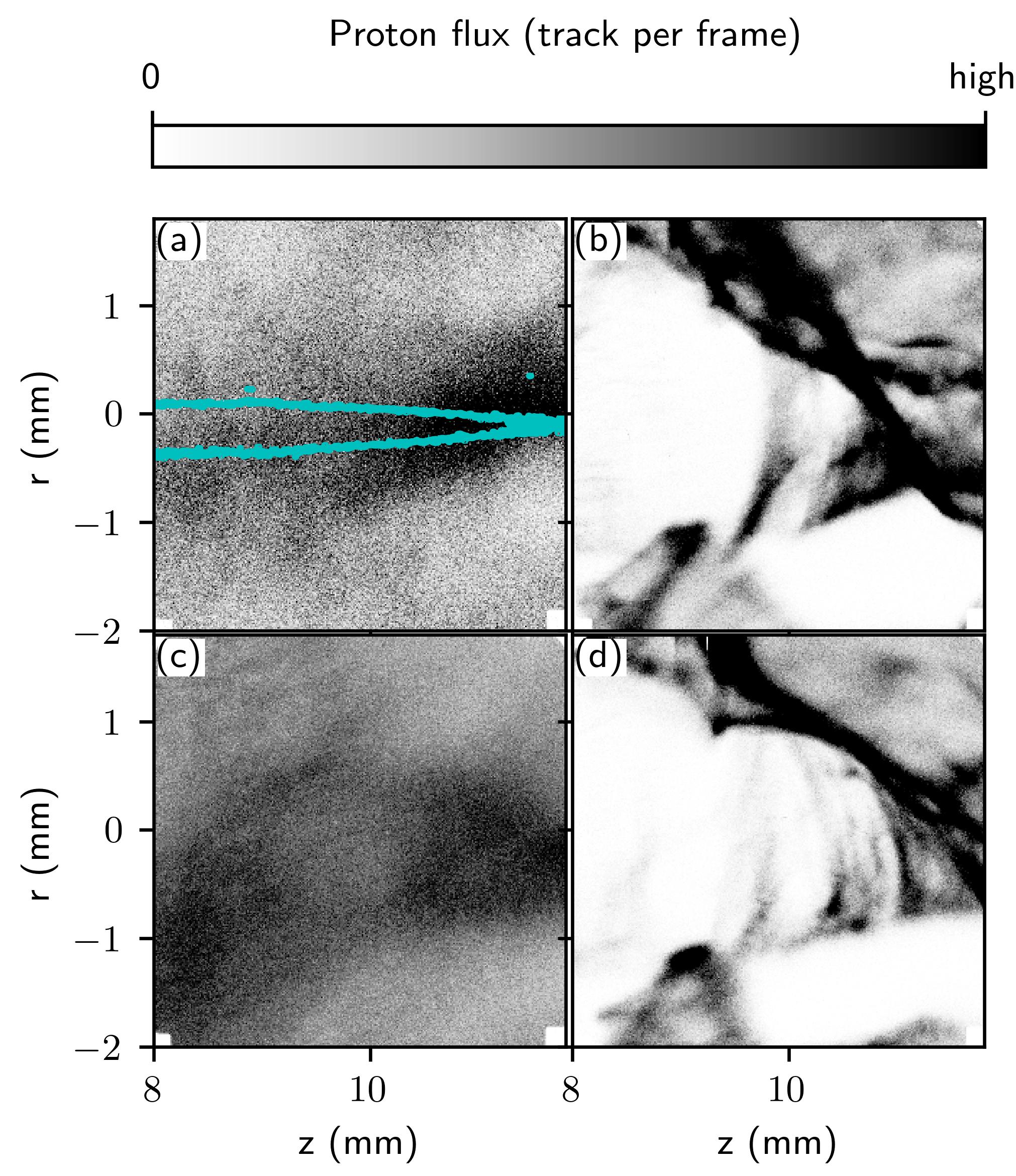}
    \caption{\SI{3}{\MeV} DD-proton radiographs captured \SI{7}{\ns} after the laser pulse, illustrating both plastic jets [(a) and (b)] and copper jets [(c) and (d)], with [(b) and (d)] and without [(a) and (c)] the externally applied magnetic field. Radiographs (b), (c), (d) exhibit a high proton fluence set to 100 tracks per frame, while (a) displays a lower fluence due to the reduced yield of the proton backlighter in this specific shot. The cyan contour delineates the core of the jet, derived from the SOP measurement, exclusively visible in (a).}
    \label{fig:prad}
\end{figure}

\begin{figure*}
    \centering
    \includegraphics{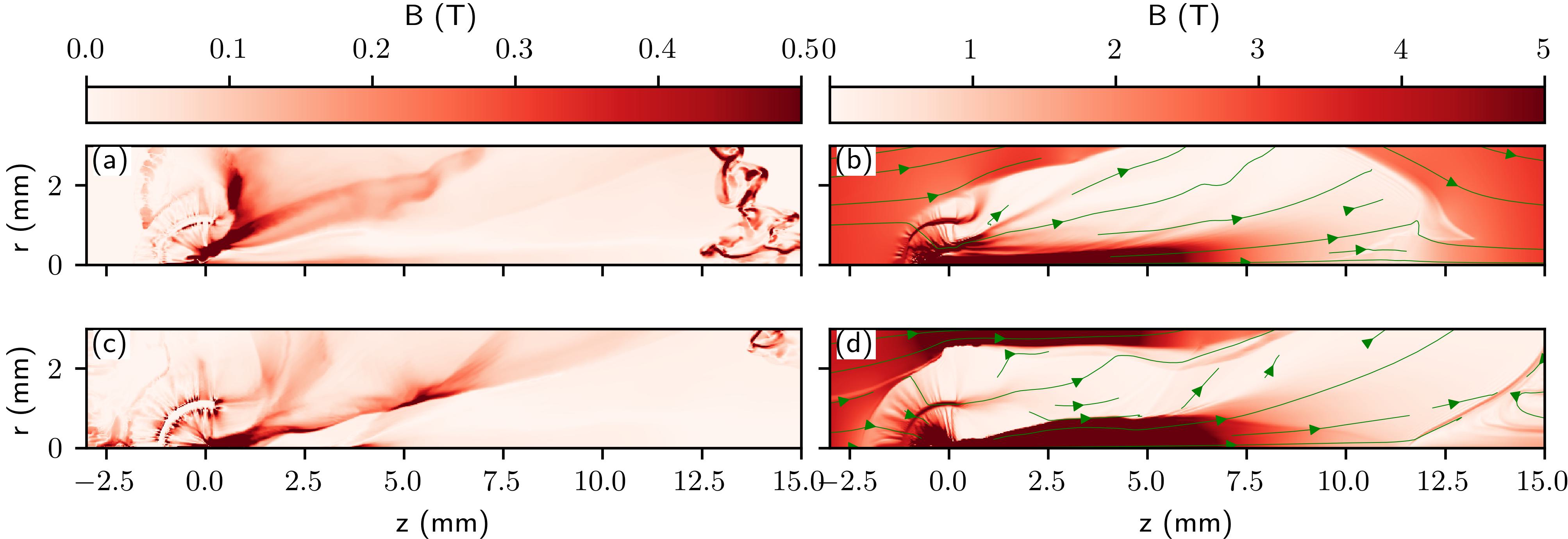}
    \caption{Simulated magnetic field norm, \SI{7}{\ns} post-laser shot, for both plastic [(a) and (b)] and copper [(c) and (d)] jets, with [(b) and (d)] and without [(a) and (c)] externally applied magnetic field. Without MIFEDS, the magnetic field is purely azimuthal. In (b) and (d), the magnetic field lines in the rz plane are displayed in green.}
    \label{fig:bsim}
\end{figure*}

The proton radiography results, obtained \SI{7}{\ns} after laser shot, are displayed in figure \ref{fig:prad}. They are obtained from the DD proton (\SI{3.02}{\MeV}), which had a higher yield than the D$^3$He reaction. Regions of low proton fluence (white) are typically associated with a high magnetic field, which drives the backlighting protons toward the lower magnetic field regions resulting in their higher fluence (dark). Here, we will distinguish the case with and without an externally applied magnetic field, the material of the jet having a lower impact on the radiographs.

First, in the absence of MIFEDS, the high flux region is concentrated in the center along the axis of symmetry. Referring to the CH case (a), the high flux zone encompasses the core of the jet (cyan line) and presents a larger flux near its tip. Here the core of the jet is extrapolated from the SOP results, by taking the jet's light emission contour and applying a constant velocity of \SI{1500}{\km\per\s}. The high flux zone is more than two times larger than the core of the jet. This corroborates the simulation results predicting the existence of an outflow surrounding the core. This interpretation is also supported by figure \ref{fig:bsim} (a and c), which shows that the magnetic field produced by the Biermann-battery effect concentrates on the edge of the jet and not in its center. One should note that the higher flux at the tip of the jet could be the result of the polar component of the produced magnetic field, which will push the proton in the direction of the axis of the jet.

In the case of an externally applied magnetic field, the morphology of the proton flux is inverted. On the symmetry axis, the bubble depleted of proton and surrounded by a high flux region is present from the direction of arrival of the jet. This bubble shows the high magnetic field resulting from the deformation of the externally applied field lines by the expanding plasma. This interpretation is corroborated by the simulations, as shown in figure \ref{fig:bsim} (b and d). As can be seen in the simulation, the core of the jet presents a high axial magnetic, due to it being pinched by the plasma. Thus, the protons are deflected toward the external part of the jet.

Both types of situations, with and without MIFEDS, show deflections on a scale larger than a millimeter. Combined with the SOP and the simulation results, these proton radiographies show the existence of a plasma surrounding the core of the jet and impacting the proton. In addition, considering the scale of the measured proton fluence as well as the simulated magnetic, one can infer that the field self-generated by the jet (Biermann battery effect) is at least an order of magnitude lower than the one applied by MIFEDS.

The X-ray results are not conclusive. While soft X-ray emission was observed near the target with a morphology that can be compared to the Bremsstrahlung emission from the simulations, the low X-ray flux and the limited size of the emission do not allow us to draw conclusions on the jet dynamics and morphology.

\section{Discussions\label{section:discussion}}
\subsection{Limitations of the simulations}
The experimental results and simulations suggest the formation of a plasma outflow is characterized by a high-velocity, high-aspect-ratio core surrounded by a low-density plasma bubble. However, the restricted scope of experimental data limits the ability to conclusively assert correspondence between simulations and experimental findings. The subsequent paragraphs address simulation limitations and highlight potential simulation artifacts.

The first approximation in the simulation comes from the laser deposition and the supposed symmetry of the experiment. While the geometry of the main target and the distribution of the laser focal spot ensure the creation of symmetric jets, the laser energy deposition is 3D by nature. Given an off-axis energy deposition in a 2D cylindrical geometry in FLASH4, the effective laser spots become a ring, whose surface, $S$, depends on the angle of the laser. Thus, the outer cone with a \SI{70}{\degree} angle has a focal spot surface greater than the real focal spot by a factor of 1.54, whereas the inner cone (\SI{70}{\degree}) has a surface lower by a factor of 1.45. As the input laser power, $P$, is the same in experiment and simulation, this results in lower (respectively higher) laser intensity, $I\propto P/S$, and thus mass flux in the coronal plasma \cite{Lindl95} coming from the high (respectively) low angle region. Considering a constant critical density for a given laser wavelength and a fluid velocity proportional to $I^{1/3}$ at the critical surface \cite{DrakeB}, the total momentum of the coronal plasma crashing in the center of the target should then be proportional to $S^{2/3}$. Given the difference in surface and the angles of the beams, this would result in a \SI{12}{\percent} increase in the simulated momentum.

\begin{figure}
    \centering
    \includegraphics{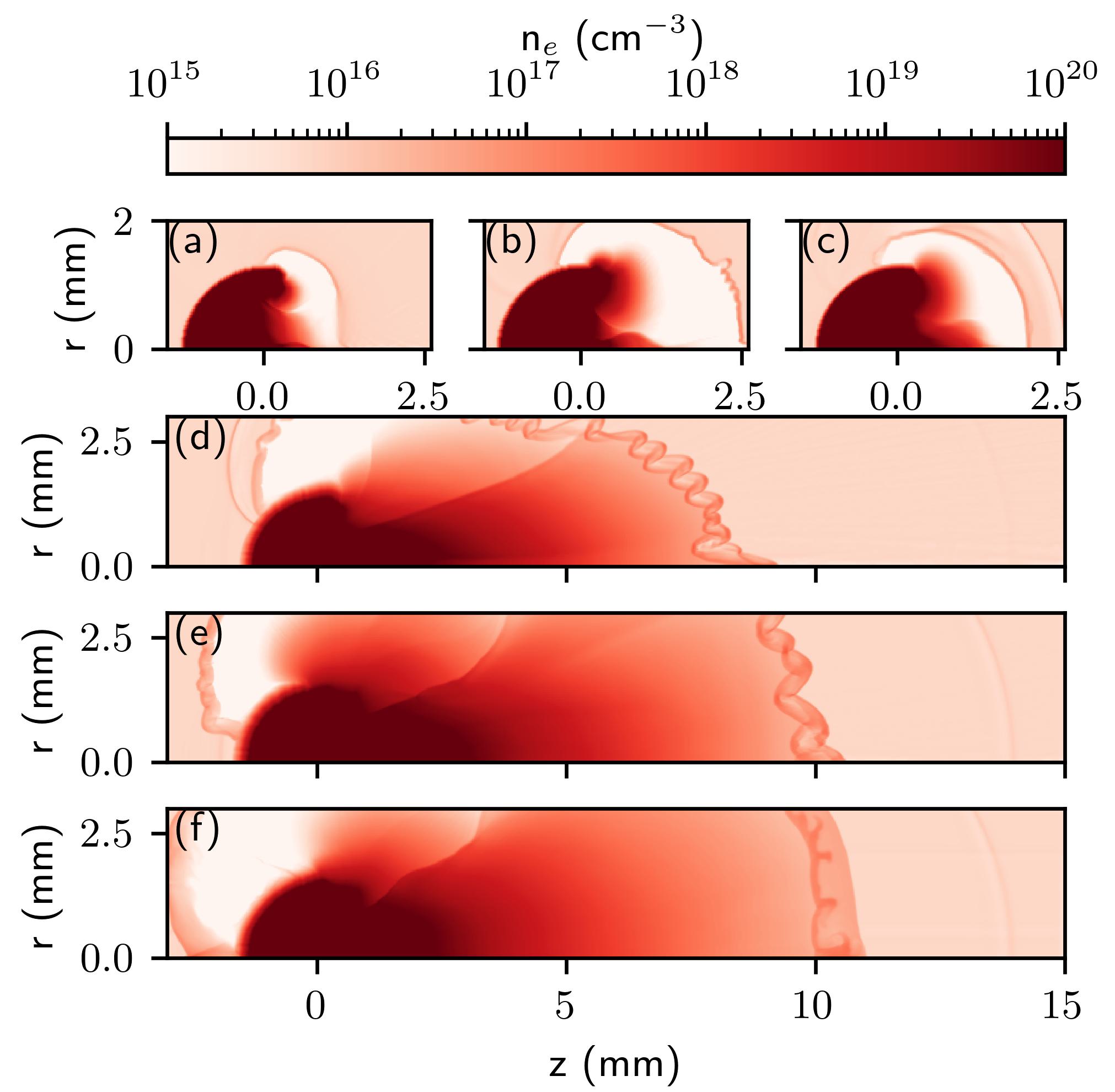}
    \caption{Comparison of simulated electron density maps according to the simulations' parameters. 2D cylindrical simulations in Conf1 (a,d) and Conf2 (b,e) are compared to a 2D cylindrical simulation initialized by a 3D simulation (c,f). Results are displayed \SI{1}{\ns} (a,b,c) and \SI{4}{\ns} (d,e,f) after the laser pulse.}
    \label{fig:2dV3d}
\end{figure}

The second approximation comes from the crashing of the coronal plasma at the target center, which is responsible for the jet creation. Here, several effects are misrepresented by the 2D cylindrical simulations. First, the use of 2D cylindrical simulation led to the most effective momentum conversion during the crash. There is no loss due to misalignment or expansion in the third dimension. While this, combined with the slightly higher momentum previously discussed, should result in a faster jet, the effect isn't obvious in simulation. Figure \ref{fig:2dV3d} shows the comparison of the electron density distribution, \SI{1}{\ns} (a,b,c) and \SI{4}{\ns} (d,e,f) after laser, according to 2D cylindrical simulations, using either only the central laser cone (a,d) or all laser cones (b,e), and a 3D Cartesian simulation, reproducing exactly the experimental setup, average over the azimuthal angle (c). The output of the 3D simulations after \SI{1}{\ns} (c) was then used as input to start a 2D cylindrical simulation to follow its later time evolution (f). As can be seen from this comparison, the geometry has only a negligible effect on the resulting jet contrary to the previous reasoning. 

At early time the lateral confinement of the jet is lower in 3D than in its 2D equivalent, but still higher than the case keeping only the central laser beams. This is also the case for the electron temperature maps, except that the central laser case also creates a high-temperature core with the same confinement as the other 2D case. This partially shows the importance of the laser distribution on the sphere for the creation of radially constrained jets. The other difference at the early time is linked to the void bubble being formed. It is smaller in the case with less laser energy as it is linked to the target radiation. It is larger in 2D compared to 3D, which might be attributed to the difference in radiation transport approximation between both cases. 

More interestingly, at late time the difference in terms of jet morphology between 2D and 3D jets while existing is negligible. So the early time difference is smoothed through the evolution rather than amplified. The case using only the central laser is slightly slower (\SI{15}{\percent}) despite receiving only one-third of the laser energy of the other cases. In terms of collimation, while neither simulation managed to correctly reproduce the observation of the SOP, the central laser case shows an apparent higher collimation over the long time despite its lower initial collimation. This is due to the radial fluctuations of the jet that are more pronounced when the outer lasers are on.

Besides these geometrical approximations, the initial convergence of the coronal plasma, which triggers the jet formation, might not be adequately modeled. Indeed, it is based on a fluid approximation in which the collision of multiple plasma plumes is perfect. While correctly simulating such interaction usually requires the use of kinetic codes, we should note that both electron and ion collision mean free paths are smaller than \SI{10}{\um} before the crash. These mean free paths were evaluated using the simulated plasma characteristics and the plasma formula \cite{NRL}. Given the simulation resolution and the initial width of the jet, these mean free paths agree with a fluid approximation.

Finally, the simulated outer part of the plasma outflow, and more precisely its interface with the vacuum species cannot be correctly simulated with FLASH4. Indeed, as FLASH uses a fluid approximation, a shock and subsequent instabilities develop at this interface. This includes an increase in density and temperature as well as an increase in the magnetic field created through the Biermann-battery effect. While these effects could exist when a fluid expands into another one, the vacuum's low density leads to a mean free path for both electron and ion larger than the jet diameter. Consequently, the fluid approximation does not hold in these regions.

In conclusion, FLASH4 simulations give a picture of the jet morphology and dynamic comparable in many parts with the one experimentally created on OMEGA. However, the simulations are expected to marginally differ from the experiments due to their 2D geometry, the associated laser deposition, and more importantly the existence of collisionless regions, that cannot be correctly simulated in a fluid approximation.

\subsection{Characterization of the jet}
\begin{table}[]
    \centering
    \begin{tabular}{|l|c|c|}
    \hline
         & YSO jets (HH 111) & Experiment \\ \hline
         Age & \num{800} year & \SI{8}{\ns} \\
         Aspect ratio & \num{>33} & \num{\sim 36} \\
         Velocity & \SIrange[range-phrase = --,range-units = single]{450}{1000}{\km\per\s} & \SI{1500}{\km\per\s} \\ \hline
         Euler & \num{e-2} -- \num{e-1} & \num{e-2} \\
         Reynolds & \num{e9} -- \num{e14} & \num{e4} \\
         Magnetic Reynolds & \num{e11} -- \num{e15} & \num{e3} \\
         P\'eclet thermal & \num{10} -- \num{e6} & 1 -- 10 \\ \hline
    \end{tabular}
    \caption{Comparison of dimensional and dimensionless number characteristic of a jet for both experiment and young stellar object (YSO). For the YSO, the HH111 case was taken as an example and the dimensional values were taken from literature \cite{Bally2016, Reipurth1989}. The experiment case uses a mix of experiment measurements and simulations for the plastic target without external magnetic fields. All intermediate values (plasma logarithm, viscosity...) needed to calculate the dimensionless number were calculated using the NRL plasma formulary.}
    \label{tab:jetcaract}
\end{table}

For this platform to be useful in the study of astrophysical jets, the conditions it produces should meet specific criteria. These criteria, usually referred to as scaling law \cite{Ryutov2000, Ryutov2001, Falize2011}, in the field of laboratory astrophysics, are summarized in two points. First, both systems (experiment and astrophysics) should share a similar geometry. And they should share similar dimensionless numbers (hydrodynamic simalarity). In table \ref{tab:jetcaract}, the dimensional and dimensionless numbers of both the experiment and HH 111, a young stellar object (YSO), are compared. While both objects differ in spatial and temporal scale, they share the same type of physical regimes. Indeed, each dimensionless number presents the same type of comparison to 1 ($\gg$, $\ll$, $\lesssim$...) showing the same kind of physical regime from the point of view of which terms can be neglected. For instance, both magnetic Reynolds numbers are much greater than 1, showing that the ideal MHD approximation can be used. However, a substantial difference exists in the order of magnitude of both Reynolds and magnetic Reynolds numbers, showing that the simulations and experiment, they modeled, don't quite reach the scale of the astrophysical jets.

\subsection{Jet angle}
When interpreting the SOP and the TS data, one of the possible explanations was the existence of an angle of the jet leading to off-axis measurements. While the orientation of the sphere might play a role, it is not believed to be a major factor. Indeed, the hole orientation might limit the plasma lateral expansion in a kind of nozzle effect and thus slightly reorient the jet, but the bulk of the jet momentum, and so the symmetry axis, comes from the distribution of the laser focal spot. Such a distribution might vary if the target has a slight spatial translation (misalignment). It should be mentioned that such misalignment is relatively low for spherical targets, bearing the utilization of MIFEDS, which led to some blind motion in the alignment procedure (target pre-alignment followed by its removal, the alignment of MIFEDS and the reinsertion of the target). Another possibility would be a non-perfect sphericity of the target surface, as the coronal plasma expands normally to it. More probably, a difference in laser energy between laser spots (\SI{\sim 7}{\percent} difference between lower and higher energy beam for each shot) could be the cause. It would lead to the non-cancellation of the radial velocity during the convergence resulting in an initial jet velocity angled from the axis.

\section{Conclusion\label{section:conclusion}}
In conclusion, a new platform to study astrophysically relevant jets was developed on OMEGA. It allows the creation of high-velocity jets, \SI{1500}{\km\per\s}, expanding over \SI{1}{\cm}, while keeping an aspect ratio of nearly 36. This platform allows the study of radiative and magnetic effects that might impact the jets. Simulations, performed using FLASH4, give some clear idea of the jet morphology and dynamics, while not perfectly matching the experimental results. Further developments and expansions of the platform are planned.

\begin{acknowledgments}
The authors acknowledge OMEGA experimental teams, which made this experiment possible, as well as all staff involved in the target production or diagnostic acquisition.

This work was supported in part by the U.S. Department of Energy NNSA MIT Center-of-Excellence under Contract DE-NA0003868, and by the U.S. Department of Energy, Office of Science and NNSA, HEDLP under Contract DE-NA0004129. 
This experiment was part of the LBS project PlasmaJet-23A. 
The 3D simulation presented in this paper was performed on the MIT-PSFC partition of the Engaging cluster at the MGHPCC facility (www.mghpcc.org) which was funded by DoE grant number DE-FG02-91-ER54109.
\end{acknowledgments}

\section*{Data Availability Statement}

The data that support the findings of this study are available from the corresponding author upon reasonable request.

\appendix

\nocite{*}
\section*{References}
\bibliography{main}
\end{document}